# SYMMETRY ENERGY CONSTRAINTS FROM GW170817 AND LABORATORY EXPERIMENTS

M.B. T$_{SANG}$[#], W.G. L$_{YNCH}$, P. D$_{ANIELEWICZ}$, C.Y. T$_{SANG}$

National Superconducting Cyclotron Laboratory and the Department of Physics and Astronomy Michigan State University, East Lansing, MI 48824 USA

## ABSTRACT

The LIGO-Virgo collaboration detection of the binary neutron-star merger event, GW170817, has expanded efforts to understand the Equation of State (EoS) of nuclear matter. These measurements provide new constraints on the overall pressure, but do not elucidate its origins, by not distinguishing the contribution to the pressure from symmetry energy which governs much of the internal structure of a neutron star. By combining the neutron star EoS extracted from the GW170817 event and the EoS of symmetric matter from nucleus-nucleus collision experiments, we extract the symmetry pressure, which is the difference in pressure between neutron and nuclear matter over the density region from $1.2\rho_0$ to $4.5\rho_0$. While the uncertainties in the symmetry pressure are large, they can be reduced with new experimental and astrophysical results.

[#]Correspondence author: tsang@nscl.msu.edu

## 1. INTRODUCTION

The Equation of State (EoS) of nuclear matter relates temperature, pressure and density of a nuclear system. It governs not only properties of nuclei and neutron stars but also the dynamics of nucleus-nucleus collisions and that of neutron-star mergers. The amount of ejected matter from the merger, which subsequently undergoes nucleosynthesis to form heavy elements up to Uranium and beyond [1-3] depends on the EoS. So does the fate of the neutron-star merger including whether: the colliding neutron stars collapse promptly into a black hole, remain a single neutron star, or form a transient neutron star that collapses later into a black hole [4]. The recent observation of a neutron-star merger event, GW170817 (GW), provides insight into the properties of nuclear matter and its equation of state (EoS) [5-8]. In this letter, we focus on how the symmetry pressure at supra-saturation densities can be extracted by combining the GW astrophysical and nuclear physics experimental constraints.

## 2. CONSTRAINTS FROM GW170817

We start with the extraction of the neutron matter EoS from the GW170817 event [5]. During the inspiral phase of a neutron-star merger, the gravitational field of each neutron star induces a tidal deformation in the other [8]. The influence of the EoS of neutron stars on the gravitational wave signal during inspiral is contained in the dimensionless constant called the tidal deformability or tidal polarizability, $\Lambda = \frac{2}{3} k_2 \left(\frac{c^2 R}{GM}\right)^5$, where $G$ is the gravitational constant, $M$ and $R$ are the mass and radius of a neutron star and $k_2$ is the dimensionless Love number [6, 8], which is also sensitive to the compactness parameter ($M/R$). As the knowledge of the mass-radius relation determines with high certainties the neutron-star matter EoS [9-12], information about EoS can be obtained from $\Lambda$.

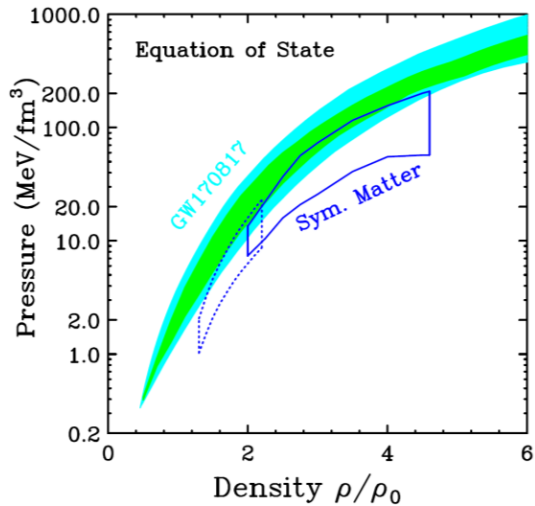

**Figure 1:** Experimental and astrophysical constraints on equation of state expressed in pressure vs. density. The green and light blue shaded regions represent the GW constraint from [6] at 50% and 90% confidence levels. Solid and dotted contours display constraints for symmetric matter from flow [22] and kaon [23, 24] measurements, respectively.

In the original GW analysis of late-stage inspiral, an upper limit of $\Lambda < 800$ was obtained assuming 1.4 solar-mass neutron stars and a low spin scenario [6]. These values of $\Lambda$ were updated in a recent analysis to $300^{+500}_{-190}$ [7]. Requiring both neutron stars to have the same EoS led to even more restrictive $\Lambda$ values of $190^{+390}_{-120}$ and $R$ values of $11.9^{+1.4}_{-1.4}$ km [6]. The corresponding 90% and 50% confidence level GW constraints on neutron matter expressed as density dependence of the pressure [6] are represented by the light blue and green shaded areas in Fig. 1, respectively.

### 3. CONSTRAINTS FROM NUCLEUS-NUCLEUS COLLISIONS

Since attaining a microscopic understanding of the EoS of dense matter constitutes an important objective of nuclear science [13], there have been ongoing experimental efforts to constrain the EoS at various densities in nuclear structure and reaction experiments [14-26] and to describe, with various success, using ab initio [27, 28], microscopic [29, 30] and phenomenological [31-33] models. However the EoS from nuclear experiments using nuclei with similar number of neutrons and protons must be extrapolated to neutron star environments where the density of neutrons greatly exceeds the density of protons. Within the parabolic approximation [34], the EoS of cold nuclear matter, expressed as the energy per nucleon of the hadronic system, $\varepsilon(\rho, \delta)$, can be divided into a symmetric nuclear-matter contribution, $\varepsilon_{SNM}(\rho, \delta=0)$, which is independent of the neutron-proton asymmetry, and a symmetry-energy term, $\varepsilon_{sym}(\rho,\delta)$ that depends on the asymmetry $\delta= (\rho_n-\rho_p)/\rho$ [35]. Here, $\rho_n$, $\rho_p$ and $\rho=\rho_n+\rho_p$ are the neutron, proton and nucleon densities, respectively. From charge symmetry, $\varepsilon_{sym}(\rho,\delta)$, must be an even function of $\delta$. Around saturation density $\varepsilon_{sym}(\rho,\delta)$ can be described by the parabolic approximation

$$\varepsilon(\rho, \delta) = \varepsilon_{SNM}(\rho, \delta=0) + S(\rho)\delta^2 + O(\delta^4) \qquad (1)$$

where $S(\rho)$ describes the density dependence term of the symmetry energy. Contributions to the EoS from known higher order terms, $O(\delta^4)$ and above, are small for $\rho<\rho_0$, less than *15%* at $2\rho_0$, but become increasingly relevant at densities greater than $2\rho_0$ [34].

The extrapolation of the EoS to neutron star environments adds $P_{sym}=\rho^2 d\varepsilon_{sym}/d\rho=\rho^2 dS(\rho)/d\rho$ to the pressure of an isospin symmetric nuclear-matter system, $P_{SNM}=\rho^2 d\varepsilon_{SNM}/d\rho$. This added pressure from symmetry energy depends strongly on the poorly constrained density dependent term $S(\rho)$ [36]. The symmetry energy influences many properties of neutron stars. Besides contributing significantly to the pressure that counters the gravitational attraction, the symmetry energy determines the proton fraction, the pressure and density of the crust-core transition and other possible phase transitions within neutron stars, and has a large impact on neutrino cooling rates by Urca and modified Urca processes [11]. Astrophysical observations do not yet provide strong constraints on the symmetry energy [37].

Above the saturation density, the neutron star pressure has significant contributions from both the symmetry energy and the symmetric matter equations of state. The pressure constraints for symmetric matter at zero temperature have been obtained from the Heavy Ion (HI) measurements of collective flow in Au+Au collisions in [22] and confirmed in [25] at densities ranging from $2\rho_0$ to $4.5\rho_0$. Similar constraints on the pressure at densities ranging from *1.2$\rho_0$* to *2.2$\rho_0$* were obtained from kaon production measurements [23, 24]. The contours enclosed by blue solid and dotted lines in Fig. 1 represent these constraints on the symmetric matter. The contours in the HI constraints are at the 68% confidence level and include the uncertainties in the measurements as well as the theoretical uncertainties in separating thermal and non-thermal constributions to the pressure and extracting the EoS from the measured data [22].

### 4. SYMMETRY PRESSURE AT HIGH DENSITY

In this section, we combine constraints from the GW170817 event [6] with laboratory constraints [22-24] to improve our understanding of the symmetry energy. As more than 90% of neutron star is composed of neutrons, bulk nuclear matter properties derived from GW170817 are close to that of the EoS of pure neutron matter, $\varepsilon_{PNM}= \varepsilon(\rho, \delta=1)$. Furthermore, the density range from $1.2\rho_0$ to $4.5\rho_0$ studied in HI experiments is comparable to that found inside neutron stars. Assuming the most probable values lie at the centers of the contours shown in Fig. 1, we can deduce the symmetry pressure, $P_{sym}=P_{PNM}-P_{SNM}=\rho^2 d(\varepsilon_{sym}(\rho))/d(\rho)$ as a function of density as shown in the right panel of Fig. 2. We note that $P_{sym}$ corresponds to the pressure contribution from the total symmetry energy and is not affected by the parabolic approximation.

For reference, we show the symmetry pressure from two commonly used bracketing assumptions for the symmetry energy within neutron-stars suggested by Prakash et al. in 1988 [38]:

$S(\rho)_{stiff}=12.7 \times (\rho/\rho_0)^{2/3}+38\times(\rho/\rho_0)^2/(1+\rho/\rho_0),$ (2)

$S(\rho)_{soft}=12.7\times(\rho/\rho_0)^{2/3}+19\times(\rho/\rho_0)^{1/2},$ (3)

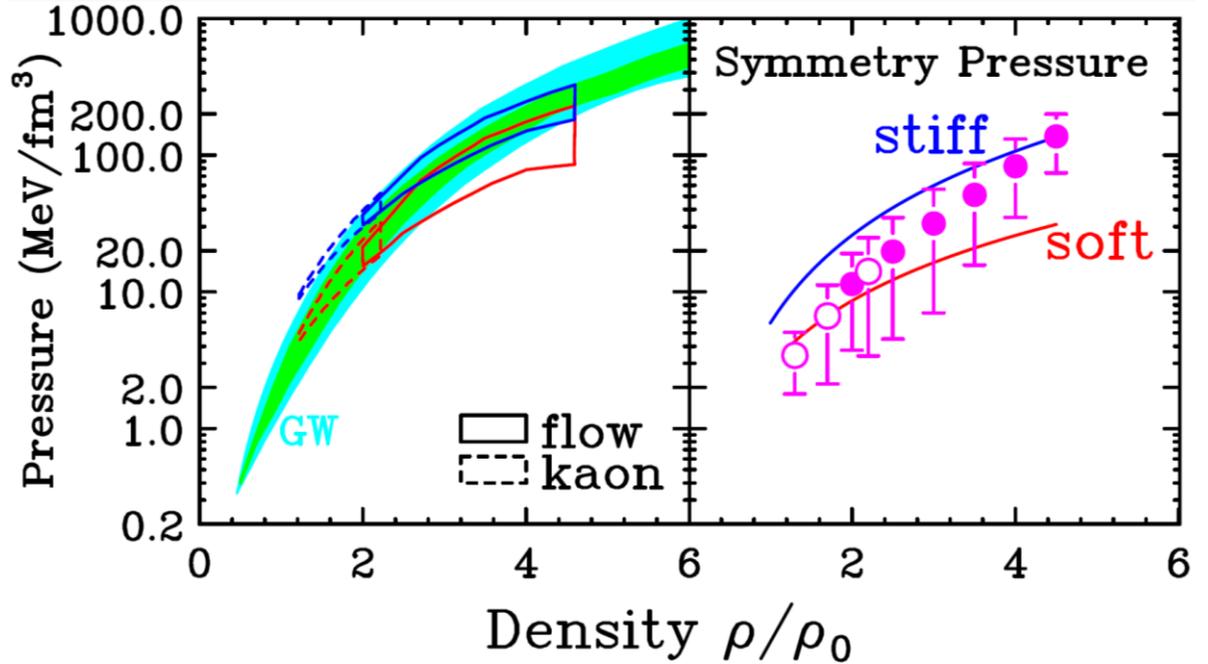

**Figure 2:** (Left panel) Laboratory and astrophysical constraints on the equation of state of neutron matter in terms of pressure vs. density. The shaded regions represents the GW constraint [6]. The upper blue and lower red contours show the expected constraints for the neutron matter EoS obtained by adding the pressure from symmetry energy functional for the stiff symmetry energy (Eq. (2)) and soft symmetry energy (Eq. (3)), respectively, to the experimentally determined pressure for symmetric matter (blue contours in Fig. 1). (Right panel) Symmetry pressure at supra-saturation densities. Circles represent 50% confidence levels values for $P_{sym}$ obtained by subtracting the most probable value of the symmetric matter pressure from the most probable value of the neutron star pressure in the left panel.

For convenience, we label Eq. (2) and Eq. (3) as "stiff" and "soft", respectively. The derived symmetry pressures are plotted as blue (stiff) and red (soft) solid curves in the right panel of Fig. 2. At low density, the most probable symmetry pressure appears closer to the "soft" symmetry energy. At $4.5\rho_0$, the data seems to agree better with the "stiff" symmetry energy. However, the uncertainties are very large.

The total pressure contours (shown in the left panel of Fig. 2) obtained by adding the stiff and soft symmetry pressures to the experimental symmetric matter pressure contours overlap with the GW constraint quite well, especially in density region below $3\rho_0$. While the constraint on $P_{sym}$ is within the expected bounds, it is clear that more accurate measurements designed to isolate the symmetry energy or symmetry pressure are needed. Experimental efforts are already underway to extract this quantity [39-41].

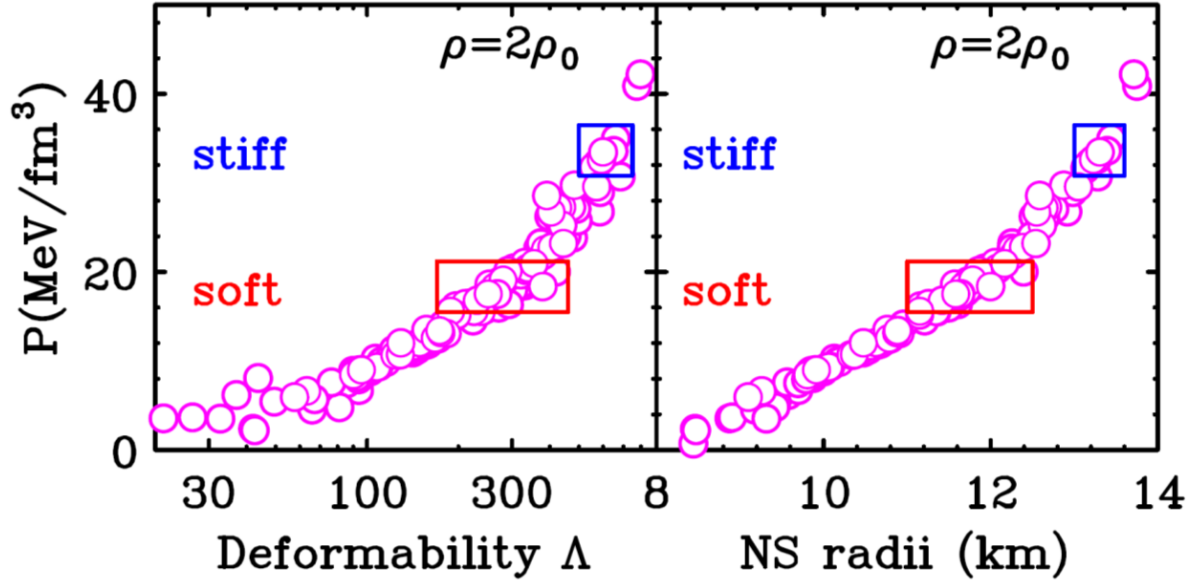

**Figure 3:** (Left panel) Correlation between pressure of a 1.4 solar mass neutron star and tidal deformability (left panel) and radii (right panel) at $2\rho_0$. The open circles are calculations based on a neutron star model that incorporates Skyrme interactions as EoS in the inner core [45]. The top (blue) and bottom (red) rectangles represent the predicted stiff and soft constraints from Ref. [22] respectively.

Using a neutron-star model that incorporates more than 200 Skyrme energy density functionals documented in [42-44] as EoS in the liquid core region of the neutron star, one can correlate the properties of neutron star with that of nuclei [45]. For example, the study shows that at twice the saturation density, there is a strong correlation between the tidal deformability as well as radii and total pressure of the nuclear system. In Fig. 3, each open circle represent the result of one Skyrme interaction parameter set. Even though the correlation illustrated in Fig. 3 applies to 1.4 solar-mass neutron stars, the trend is the same for different mass neutron stars. Thus, an experimental determination of the total and symmetry pressure at $2\rho_0$ would, in turn, set constraints on $\Lambda$ and $R$. For illustration, adopting Equation (2) or (3) for $S(\rho)$ with the same experimental and theoretical uncertainties in [22], one would obtain a hypothetical experimental pressure represented by the upper (blue) or lower (red) rectangules in Fig. 3, respectively. An hypothetical experimental value of the pressure between *15.5-21.2 MeV/fm³* corresponding to the soft symmetry dependence of Eq. (3) would be consistent with the tidal deformability and radii values in the range of $\Lambda$~170–460 and R~11-12.5 km (red rectangles), respectively. Similarly, with a measured pressure between *30.8-36.5 MeV/fm³*) that is consistent with the stiff symmetry dependence of Eq. (2), one would obtain $\Lambda$ ~ *500–750* and *R~13-13.6 km* (blue rectangles). Here, we use the Skyrme interactions in the neutron-star model described in ref. [45] for illustration and discussion purpose. The conclusion, i.e. specific $\Lambda$ and $R$ values extracted depend on the different forms of EoS used in the neutron-star model. Looking forward, the expected improvement in heavy-ion and in the gravitational wave constraints will allow us to understand what these combined constraints imply for the microscopic nature of strongly interacting matter within the neutron-star environment.

5. SUMMARY

In summary, by combining the gravitational wave and the symmetric matter EoS from heavy ion measurements obtained 16 years apart, we deduced the most probable values for the symmetry pressure between 1.2 to 4.5 times the saturation density. Amazingly, the extracted symmetry pressures are consistent with the symmetry energy parameterization used in a dense matter model by Prakash et al in 1988. The current work suggests that laboratory studies of nuclear matter around twice the saturation-density will yield observables that can be directly related to neutron star radii and deformability. As the precision of GW constraints on neutron matter and heavy ion constraints on symmetry energy and symmetric matter improve, more stringent constraints on the symmetry energy are expected.

6. Acknowledgement

This work was partly supported by the US National Science Foundation under Grant PHY-1565546 and by the U.S. Department of Energy (Office of Science) under Grants DE-SC0014530, DE-NA0002923 and DE-SC001920.